# A study of uncompensated latency in ADS-B reports

Michelle George
Arcon Corporation
michelle@arcon.com

Saurav Tuladhar
Arcon Corporation
saurav@arcon.com

Siva Sivananthan
Arcon Corporation
siva@arcon.com

**Abstract—** The total latency (TL) in ADS-B is the difference between the true TOA of the original position measured by the onboard GPS and the TOA in the ADS-B report at the GS. We processed the ADS-B reports extracted from live-data recorded at the Chicago to produce the results in this report. We observed that 99% of aircraft in our data set were broadcasting in non-UTC coupled mode. For these aircraft, STARS exclusively used ADS-B reports to update tracks when available. The calculated ULs for the non-UTC coupled ADS-B reports were well within the budget (-200 ms to +400 ms) allowed by the DO-260B MOPS. Moreover, the mean latencies from both methods were less than ±50 ms. The latencies obtained from the MTPES method were observed to be in close agreement with the means from the ATPE method. About 1% of aircraft were broadcasting in UTC coupled mode. The reports from these aircraft were identified to have a non-compliant link version, causing STARS to reject these reports for tracking. Moreover, the reported position and TOA were found to be unsynchronized, which conflicts with the MOPS requirement for UTC coupled operation.

## I. INTRODUCTION

Automatic Dependent Surveillance – Broadcast (ADS-B) is the key surveillance technology in the FAA's NextGen project [1]. ADS-B equipped aircraft automatically broadcast aircraft identity and state information (position and velocity) at periodic intervals. The ADS-B messages broadcast from an aircraft are received by the ground stations (GS) and other ADS-B equipped aircraft within the broadcast range. The ADS-B messages received at the GS are delivered to STARS for aircraft tracking. The ADS-B messages received by other ADS-B aircraft are used by them for collision avoidance and maneuver selections.

An onboard ADS-B transmitter depends on an onboard GPS source for aircraft position and velocity information. The GPS derived position information is known to provide accuracy ranging in meters (Table 2-70 [2]). Similarly, the GPS derived velocity information is also accurate within 10 m/s (Table A-5 [2], [3]). The onboard ADS-B transmitter processes the data from the GPS source and assembles the message for broadcast. There are two physical layer variants for transmitting ADS-B messages: the Universal Access Transceiver (UAT) and the 1090 MHz-extended squitter (1090-ES). This report focuses only on the ADS-B version that uses the 1090-ES.

The broadcasted ADS-B message contains aircraft ICAO address, aircraft position and velocity, and the associated reliability and uncertainty category indicators (NIC, SIL, NAC, NUC). This information is broadcasted in two position messages and two velocity messages per second. Notably, the time of applicability (TOA) of the position and velocity information is not included in the broadcasted message. Instead, the GS assigns a TOA timestamp to the received ADS-B messages based on the time of message reception (TOMR). A report assembly function at the GS assembles the ADS-B message and TOA timestamp into an ASTERIX CAT033 report format [4]. The ASTERIX CAT-33 reports are delivered to STARS for target tracking. The ASTERIX CAT-33 reports will be referred to as ADS-B reports in this report.

STARS uses the TOA information in the ADS-B report as the valid time of the position for tracking application. However, the true TOA of the reported aircraft's position and velocity is not necessarily the same as the GS assigned TOA due to various latencies incurred in the ADS-B system (transmitter, propagation channel, ground station). Since STARS uses the ADS-B report TOA for tracking, it assumes the TOA is valid. With this assumption, the impact of transmission latency can be seen as the state information in the report being 'delayed' compared to the TOA. This delayed state information is less accurate for the assigned TOA than what is reported by the associated navigation accuracy category - position (NACp). The sources of latency in ADS-B are detailed in Appendix U [2]. This report presents results of ARCON's work to quantize the latency in ADS-B by mining live recorded data.

The onboard ADS-B broadcasting system compensates for the transmission latency by adjusting the position information in the broadcast message. The latency compensation can account for only a part of the total latency; the latency that remains is called uncompensated latency. The uncompensated latency (UL) induced position error is a known issue with ADS-B reports [5], [6], [7] [8]. ARCON was tasked with mining large amounts of ADS-B data recorded at a STARS site to identify UL in ADS-B reports. The data mining process involves applying analytical tools and techniques to discover patterns and characteristics in large datasets. As a part of the data mining process we performed following tasks:

- Extracted ADS-B reports and track data from STARS 8001-LAN messages recorded at the Chicago STARS site.

- Used the track data to generate a pseudo-truth track. This pseudo-truth track provides a reference position and time necessary to identify UL in ADS-B reports.

- Computed UL in ADS-B reports using the pseudo-truth and ADS-B reported velocity. We computed UL in ADS-B reports from 40 aircraft and analyzed their characteristics.

- Identified anomalies in ADS-B reports from aircraft in UTC coupled mode.

This report presents the pseudo-truth generation method and the UL calculation methods used in this task and also discusses the results obtained after analyzing the UL in the ADS-B reports.

This report is organized as follows: Section **Error! Reference source not found.** presents an overview of the time management approach in the ADS-B system and the inherent latency issue in ADS-B reports. This section also discusses the impact of latency on position accuracy. Section **Error! Reference source not found.** discusses the UL calculation approach used to generate results presented in this report. The pseudo-truth generation method and two variants of latency calculation algorithms are presented in detail. This section also gives an overview of the live-data recorded at the Chicago STARS site. Section **Error! Reference source not found.** presents the results from UL analysis of the ADS-B reports in the Chicago live-data. We also present our concluding remarks in this section.

## II. TIME MANAGEMENT AND LATENCY IN ADS-B

ADS-B messages received by the GS do not contain the TOA associated with the position or velocity information in the message. The GS assigns the TOA timestamps to the received ADS-B message using two different methods depending on the onboard ADS-B transmitter configuration. If the ADS-B transmitter is synchronized to a Coordinated Universal Time (UTC) time source such as the GPS, the transmitter equipment is said to be in "UTC Coupled" mode. In the UTC coupled mode, the position and velocity in an ADS-B message are synchronized to a precise moment in UTC time. In this case, the GS report assembly function calculates the TOA as the nearest even or odd 200ms UTC epoch, depending upon the type of ADS-B message [4]. A UTC time synchronized configuration is indicated by the TIME (T) bit set to 1 in the ADS-B message. Only about 1% of aircraft in the Chicago live-data recording have UTC coupled ADS-B transmitters.

On the other hand, if the onboard ADS-B transmitter is not synchronized to a UTC time source, the transmitter equipment is said to be in "non-UTC coupled" mode. In the non-UTC coupled mode, the position and velocity in the ADS-B message is not synchronized to UTC time. In this case, the only precise timestamp available at the GS is the TOMR. The GS report assembly function calculates the TOA by rounding the TOMR to the nearest 1/128$_{th}$ second [4]. A non-UTC coupled configuration is indicated by the TIME (T) bit set to 0 in the ADS-B message. About 99% of the aircraft in the Chicago live-data recording have non-UTC coupled ADS-B transmitters.

In the UTC coupled case, the transmission latency can be essentially eliminated as the GS is able to assign a TOA to a received message from a synchronized UTC time source. In the non-UTC coupled case, latency is introduced due to the processing delays in the ADS-B transmitter, propagation delay in the channel, and processing delay in the ground station (Appendix U [2]). The latency can be partially compensated for by propagating the position information to a chosen time of transmission. But the uncompensated portion of the total latency remains as an issue. The following section describes the latency issue in ADS-B and discusses the compensated and uncompensated components of the total latency.

### A. Latency Definition

Total latency (TL) is the difference between the true TOA of the original position measured by the GPS and the TOA in the ADS-B report. The ADS-B receivers take the TOMR as the TOA for the received position. There can be significant latency error in a position message due to various processing delays between when the measurement is made and when the measurement is received. The onboard ADS-B transmitter propagates the original position measurements to reduce this latency before transmitting. This is done by selecting a future time that it intends to transmit the position measurement. Using the measured velocity, the initial position measurement is propagated to the chosen transmission time. This propagated position is broadcast in the ADS-B message. Since the broadcast messages are assumed to be received instantaneously the GS assume that TOMR is equal to TOA of the position.

However, there can be considerable latency left unaddressed by this compensation process. For example, the transmitter may choose an intended transmission time that is later than or earlier than the actual transmission time. Additionally, there can be processing delays in the avionics before this compensation takes place, or there can be a delay when the message propagates through the channel. Also, there can be delays in the GS between the time a message is received and the time at which it is timestamped. The portion of TL that is not compensated by this propagation operation is the uncompensated latency (UL).

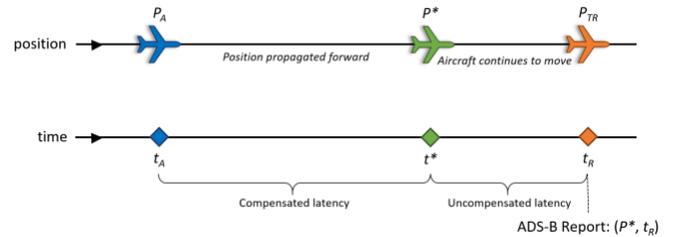

Fig. 1. Definition of latency in ADS-B report.

Fig. 1. illustrates the two components of the total latency incurred in the ADS-B system for a single position measurement. Fig. 1. assumes that the position measurement has no errors. The initial position measurement and the corresponding time are ($P_A$, $t_A$). This measured position information cannot be transmitted instantly due to necessary

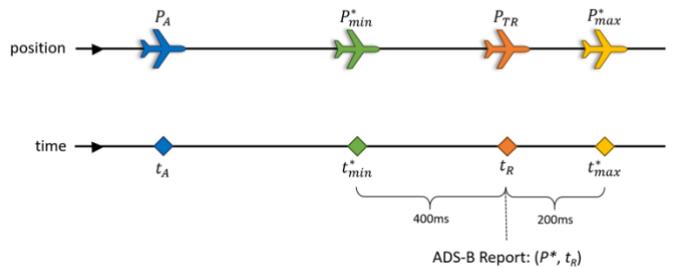

Fig. 2. Over-compensation vs. under-compensation resulting in positive or negative latencies.

onboard processing. Time $t^*$ is the selected time to transmit the ADS-B message. In order to compensate for the delay between $t_A$ and $t^*$, the position $P_A$ is propagated using the delay $(t^* - t_A)$ and the measured velocity. Hence, $t^*$ is the true TOA for the propagated position $P^*$. This position $P^*$ is inserted in the ADS-B message to be transmitted. However, due to additional onboard processing delay the ADS-B message is actually broadcast at time $t_R$. The delay $t_R - t^*$ is defined as the uncompensated latency (UL). Assuming negligible propagation and processing delay, the timestamp applied by the GS to the received ADS-B message can be taken as $t_R$. However, at time $t_R$ the aircraft has continued to move along-track to a new position $P_{TR}$. Hence, the position $P^*$ received at the GS is a delayed position for the associated time $t_R$.

There are three sources of latency in the onboard ADS-B equipment described in Appendix U of [2]. Each of these three latency sources meets specific timing requirements. These requirements on each latency source lead to an allowed UL budget of -200 to 400 ms. Positive and negative latencies result due to under-compensation and over-compensation respectively when propagating the aircraft position before transmission. Fig. 2. demonstrates the propagated positions corresponding to the extrema of the UL budget. In this figure, the orange diamond marker denotes the actual broadcast time $t_R$. The green aircraft marker denotes the position ($P^*_{min}$) propagated to time $t^*_{min}$ which is 400ms behind $t_R$. Using the position $P^*_{min}$ in the ADS-B message will incur a UL of $t_R - t^*_{min} = 400ms$. Similarly, the yellow aircraft marker denotes the position ($P^*_{max}$) propagated to time $t^*_{max}$ which is 200ms ahead of $t_R$. Again, using the position $P^*_{max}$ in the ADS-B message will incur a negative UL of $t_R - t^*_{max} = -200ms$. The DO-260B MOPS requires the ADS-B avionics to keep the UL within the allowed range.

Following the position and timing definitions in Fig. 2., we aim to calculate the UL for a given ADS-B report. If the true TOA $(t^*)$ corresponding to ADS-B reported position $(P^*)$ is known, the UL can be calculated directly as $UL = t_R - t^*$.

Alternatively, if we know the true position $P_{TR}$ corresponding to a reported timestamp $t_R$, the UL can be approximately calculated as:

$$UL = \frac{|P_{TR} - P^*|}{|v|} \quad \text{(II.1)}$$

where $|v|$ is the aircraft's true speed. The numerator in Eq. (II.1) is the error between the reported position and the true position corresponding to $t_R$. This definition assumes that the aircraft is traveling in a straight-line trajectory with a constant speed.

In practice, the true position $P_{TR}$ or the true TOA $(t^*)$ are both unknown and the above definitions cannot be used directly to calculate UL.

### B. Impact of the Uncompensated Latency

STARS assumes the ADS-B reported TOA $(t_R)$ is the valid time for the reported position $(P^*)$. As discussed above, the reported position is inaccurate for this TOA. The true position at the reported TOA $(P_{TR})$ equals the true position vector corresponding to $P^*$ plus error due to the aircraft moving along its trajectory during the onboard processing. This additional error is a result of the UL. Hence, the true position at the reported TOA is

$$P_{TR} = P^*_T + e_{UL} \quad \text{(II.2)}$$

where $e_{UL}$ is the error vector due to UL.

The reported position $P^*$ is

$$P^* = P^*_T + e_n \quad \text{(II.3)}$$

where $e_n$ is the measurement noise and $P^*_T$ is the truth corresponding to the reported position. The distribution of measurement noise $e_n$ is characterized by the NACp value in the ADS-B report. In the above equations, the position and error variables represent vector quantities.

Now, using Eq. (II.2) and Eq. (II.3), the numerator of Eq. (II.1) can be expressed as

$$|P_{TR} - P^*| = |e_{UL} - e_n| \quad \text{(II.4)}$$

Eq. (II.4) defines the error in the ADS-B reported position. The total error is comprised of measurement noise $e_n$ and the error introduced due to UL, $e_{UL}$. Observe that when there is no UL, the only source of error is the measurement error. But a high UL results in large position error, even when NACp is high and $e_n$ is small. Consequently, the reported NACp does not reflect the true accuracy of the reported position. Currently, STARS hard-limits ADS-B reported NACp to 8 to account for UL induced error in the position report.

*a) Along-track and cross-track errors*

Assuming a straight-line trajectory, the UL induced position error falls along the track. The measurement noise can be decomposed into along-track and cross-track components as $e_n = e_{nAT}i + e_{nXT}j$, where $i$ is a unit vector in along-track direction and $j$ is a unit vector along the cross track direction. Using this decomposition and the fact that $e_{UL}$ is in along track direction, Eq. (II.4) can be expressed as

$$\begin{aligned}|P_{TR} - P^*| &= |e_{UL}i - e_{nAT}i - e_{nXT}j| \\ &= |e_{AT}i - e_{nXT}j|,\end{aligned} \quad \text{(II.5)}$$

where $e_{AT} = e_{UL} + e_{nAT}$ is the total along-track error vector. Hence, in a true straight-line trajectory the along-track error is composed of the UL induced error and the along-track component of the measurement noise. Hence, if we assume the true velocity along the straight-line trajectory is $v$, an improved estimate of the UL can be calculated as

$$UL = \frac{|e_{AT}|}{|v|} \quad \text{(II.6)}$$

This definition of UL contains the actual UL plus the offset due to the along-track measurement noise. The impact of the cross-track measurement noise is eliminated. The following sections discuss simulation results which demonstrates the impact of measurement noise in the calculated UL

### III. LATENCY ESTIMATION

Section II.B suggests that we can use Eq. (II.6) to estimate UL. Eq. (II.6) assumes that the true position at TOA (PTR) is known and the along-track error can be calculated. In practice however, the true trajectory of the aircraft is unknown.

Identify applicable funding agency here. If none, delete this text box.

Specifically, in this task we aim to estimate UL from ADS-B reports recorded at a live STARS site. Hence, we strive to find an approximation to the true trajectory from the recorded data for the aircraft broadcasting ADS-B messages. The true trajectory is approximated with a pseudo-truth which is generated using filtered-track data available from STARS. The pseudo-truth track is denoted as function in time $P(t) = [x(t), y(t)]$. Once the pseudo-truth track function is known it can be evaluated at the desired TOA to determine a reference position. Alternatively, an inverse operation produces reference TOA at a desired position.

The data recording facility at a STARS site records the 8001-LAN messages and the ASTERIX CAT033 reports along with other information. The 8001-LAN message contains track data of all the aircraft being tracked. The track data contains estimated horizontal position, velocity, and timestamp information. In this task, we use track data to generate pseudo-truth for aircraft trajectories available in the recording. The ASTERIX CAT033 reports in the recording contain the ADS-B data received at the GS. These ADS-B reports together with the pseudo-truth are used to calculate the UL.

The following section describes the spline fit based pseudo-truth track generation method. All results presented in this report are based on the pseudo-truth trajectory generated using one of the methods described below.

### A. Spline Fit Pseudo-Truth Track

A polynomial spline fit operating on the track data can generate the pseudo-truth track function that approximates the true track. A spline fit produces a function that represents a series of positions (points in x-y space) using a series of polynomials as a function of time. The choice of using a polynomial spline fit to derive the pseudo-truth track is motivated by the fact that aircraft motion in the x and y dimension is a polynomial function in time [9]. Also, multiple authors have published literature discussing spline-fit based smoothing of maneuvering trajectories [10] [11].

In this task, we use a cubic (k=3) polynomial spline to fit each track. The spline-fit was performed on x-positions (t, x) and y-positions (t, y) to produce spline representations of each dimension with respect to time. The first order term in the cubic polynomial represents a constant velocity motion and the second and the third order terms capture acceleration and jerk in the track data. Using the cubic polynomial spline to fit constant velocity straight-line track risks overfitting the spline to noisy position data. But we can specify a smoothing condition to minimize the risk of overfitting.

The smoothing parameter s defines the smoothness of the spline fit. The amount of smoothness is determined by satisfying the condition $\sum(x - g)^2 \leq s$, where g is the spline fit of (t, x). Setting s = 0 produces an interpolating spline whereas choosing s > 0 enforces smoothness over closeness of fit to the input data. Selecting an appropriate positive smoothing condition generates a smoothed spline to represent the pseudo-true position. We used an iterative approach to select an appropriate smoothing condition for each track selected for analysis.

*1) Iterative Smoothing*

First, we aim to set some criteria that will define the smoothness of the pseudo-truth track. As described above, the spline fit will result in a cubic polynomial that defines the aircraft kinematics. The second derivative of this curve represents the acceleration. Commercial aircraft fly constant velocity straight-line trajectories for most of the flight duration. We can safely assume that a commercial aircraft's acceleration will remain low or near-zero for the duration of the track.

We can retrieve approximate acceleration information for a track based on the ADS-B reported velocity. If we have the knowledge of the expected acceleration for a given track, we can leverage this information to set bounds on our pseudo-truth. The derivative of reported ADS-B x-velocity and y-velocity is used to establish the maximum and minimum allowable acceleration of the pseudo-truth track.

To begin smoothing, we start with an initial value s = 0. We evaluate the resulting spline fit track smoothness by calculating its acceleration. The value of s is incrementally increased until the pseudo-truth acceleration lies within the acceleration limits.

After each spline fit, we calculate the sum of squared residuals of the spline approximation. This sum is a measure of the closeness of the spline fit to the original input data. Because data in the x-dimension and y-dimension are fitted separately, we obtain two residual sums for each track.

In this report we present two variants of UL calculation methods. Section III.B presents a method that uses the along-track position error and ADS-B reported speed to calculate UL as defined in Eq. (II.6). This method is based on calculating the latency in each individual reported position. Instead of examining a set of individual latencies from ADS-B points, we can also calculate the overall latency by assigning a single UL to all ADS-B reports so as to satisfy an optimization criterion. Section III.C presents this second method which calculates a single latency for each track.

### B. Along-Track Position Error Method

The along-track position error (ATPE) method uses Eq. (II.6) to estimate the UL in ADS-B reports. This method begins by calculating the along-track error for each ADS-B reported position. Fig.3. illustrates the along-track error calculation for a single ADS-B report. In this figure, the blue marker represents a single ADS-B reported position ($P^*$). The red diamond marker denotes the corresponding reference position $P(t_R)$ on the pseudo-truth track. This reference position is obtained by evaluating the pseudo-truth track function at the ADS-B report time $t_R$. The green vector represents the along-track error vector.

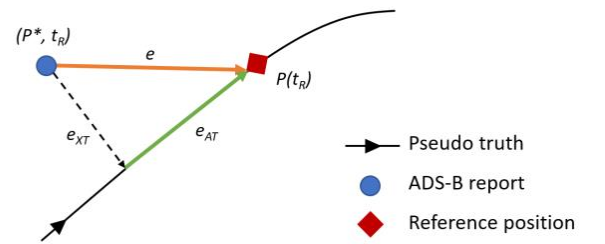

Fig. 3. Along-track position error-based UL calculation for a single ADS-B position.

For each ADS-B report, we calculate the UL in following steps:

Evaluate the pseudo-truth function at the ADS-B report TOA, obtaining the pseudo-true position P(tR). This serves as our reference position.

Calculate the position error vector $e = P(t_R) - P^*$

Project (vector projection) the total position error vector e onto the along-track unit vector to determine the along-track position error eAT.

Calculate the uncompensated latency using Eq. (II.6),:

$$UL = \frac{|e_{AT}|}{|v|}$$

where $|v|$ is ADS-B reported speed, which is the along-track speed.

We calculate a latency estimate for each individual ADS-B point on the track. The overall distribution of the track latencies can be characterized by building a histogram of the calculated delays. This provides insight into the distribution of latencies for that track. Results from specific tracks using these methods are examined in more detail below.

## C. Minimization of Position Errors Approach

In this method we calculate a single latency value for an entire track that aims to minimize ADS-B position error. This method assumes that the probability density of ADS-B track latency can be modeled as a uniform distribution for the duration of the track. The position error of the nth ADS-B report can be written as $\|P(t_{Rn} - \Delta T) - P_n^*\|$ where $\Delta T$ is the desired latency.

As explained earlier, this error is a combination of error due to UL and error due to jitter in the measurement. We collect the square of this error for multiple points (N) and try to find $\Delta T$ that will eliminate the error due to UL and leave only the jitter component in the total of the squares. If the NACp remained constant in the track, then the remaining error should satisfy the EPU corresponding to that NACp. Thus, this calculation can be formulated as a constrained minimization problem where we perform minimization of total position error squares (MTPES) with respect to $\Delta T$ such that the remaining error satisfy the EPU corresponding to that NACp.

We explored two variants of this method to calculate the latency. The first method calculates the latency such that 95% of the $N$ position errors satisfy the ADS-B reported EPU bound. In our exploratory analysis, we observed that this method produces latency values in cases where the reported NACp is 9 and lower. When the reported NACp > 9, the EPU bound is less than 0.005 nmi. In these cases, the 95% EPU satisfaction constraint becomes impossible to meet and no solution is obtained.

The second method, MTPES, calculates the latency that minimizes the total of position-error squares. This minimization problem can be expressed as a solution to the optimization problem,

$$UL = \arg\min_{\Delta T} \sum_{n=1}^{N} \|P(t_{Rn} - \Delta T) - P_n^*\|^2,$$

where $P(t)$ is the pseudo-truth function, $(P_n^*, t_{Rn})$ is the position and TOA of the nth ADS-B report, and $N$ is the number of ADS-B reports available.

In practice, this method will be applied offline to ADS-B reports collected from a single aircraft. The latency solution to the above equation can be used to adjust the TOA in ADS-B messages from that aircraft, thus minimizing the UL induced position errors. The latency results to be discussed in Section IV.A shows that the latency calculated using the MTPES method is consistently in agreement with the mean of the histogram obtained from the ATPE method.

## IV. RESULTS AND DISCUSSION

In this task we analyzed the ADS-B reports extracted from live-data recorded at the Chicago (ORD) STARS site. The data was recorded on June 8th, 2018 and spans a 12-hour period from 13:00 to 01:00 hours UTC time. We used in-house utilities to extract the ADS-B reports and track data from the recording file.

Examining the track data indicated that the 12-hour period recording contained tracks from 776 unique ICAO addresses. We noticed that certain ICAO address had multiple tracks occurring at different times of the day. Before proceeding with UL calculation and analysis, we filtered the track data to keep data only from ICAO addresses that meet the following criteria:

- ADS-B reports must use the 1090-ES link
- ADS-B reports have NACp between 8 and 11
- ADS-B reports must be non-UTC coupled
- Have at least two unique tracks

We chose to have at least two unique tracks per ICAO address to investigate if the ULs from the same aircraft change over time. Using these filtering criteria narrowed the data to include a set of 40 unique ICAO addresses spanning 82 individual tracks. Fig.4. shows the geodetic positions of the selected 82 tracks. The star marker indicates the ADS-B ground station position. This ground station is located inside the Chicago O'Hare International Airport (ORD) premises.

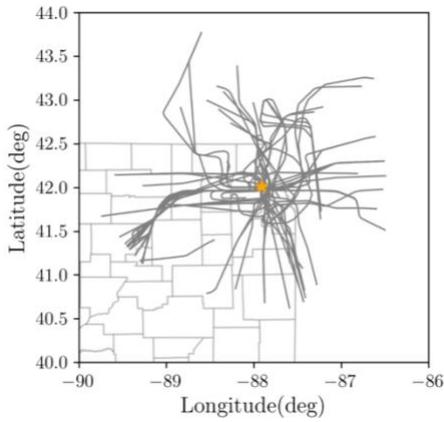

Fig. 4. Figure 1 Selected aircraft tracks for latency analysis.

*A. Latency Results*

We calculated latency in every individual ADS-B report from every track as described in the previous section. These latencies were calculated using the pseudo-truth based methods described in Section III. Fig.5. shows the distribution of estimated latencies and standard deviations for all the 82 tracks belonging to the selected 40 aircraft. In Fig.5., the y-axis represents the latency in milliseconds and the x-axis represents track indices from 0 to 81. For each track, the mean latency is indicated with a square or a diamond marker. The vertical lines with circular markers at the endpoints represent one standard deviation interval on either side of the mean. Tracks belonging to the same aircraft are grouped by color and marker style. For instance, the first two vertical lines on the left belong to two tracks from a single aircraft.

In **Error! Reference source not found.**, observe that the estimated mean latencies for all tracks lie within the span of -60ms to 60ms, well within the bound specified in Appendix U of [2]. Additionally, the standard deviation bars demonstrate that 68% of all individual estimated latencies fall within the range of -105ms to 105ms.

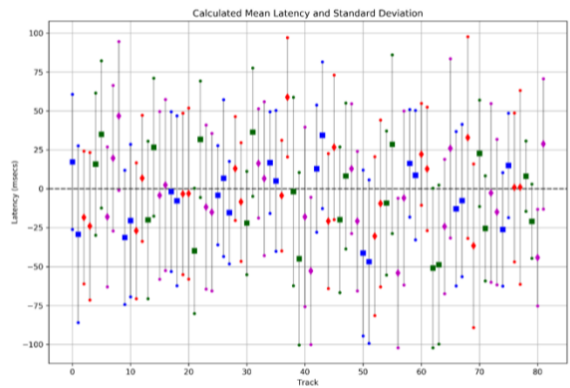

Fig. 5. Mean latency and standard deviations for all 82 tracks in the dataset.

In Fig. 5., we can observe positive and negative mean latency exhibited across aircrafts. For example, the aircraft at track indices 60 and 61, shown with red diamond markers, demonstrates a consistent positive mean latency. The aircraft immediately following, shown with green square markers at indices 62 and 63, has a clear negative mean latency close to -50 ms for both tracks. Finally, the aircraft with purple diamond markers at indices 64 and 65 demonstrates a skew in mean latencies; -24.316 ms for one track and 23.258 ms for the other. These cases represent specific notable behaviors exhibited throughout the entire dataset of 40 aircraft.

*B. Anomalies in UTC Coupled ADS-B reports*

About 1% of aircraft (10 out of 776) in the Chicago STARS site dataset were broadcasting ADS-B messages with the UTC Coupled flag set to 1. The UTC coupled flag set to 1 indicates that the broadcasting aircraft's avionics are synchronized to a UTC time source. When a UTC coupled ADS-B message is received, the GS assigns a TOA that is the nearest 200 ms UTC epoch. Examining the UTC coupled ADS-B reports we observed:

TOAs are resolved at the GS to a 200 ms UTC epoch. Fig.6. shows the fractional portion of the TOA in the ADS-B reports received from aircraft A637E1. The y-axis represents the fractional second in TOA and the x-axis represents the UTC Time of Day of the ADS-B report. Each blue dot corresponds to a single ADS-B report. It is evident that the fractional second in each TOA is exclusively a multiple of 200 ms. Identical results were observed for the other nine UTC coupled aircraft.

ADS-B reports of UTC coupled aircraft were not used for track updates in STARS. The track is updated exclusively using radar reports. Further investigation revealed that the ADS-B reports from these 10 aircraft were using link version 1 which is non-compliant with the DO-260B standard. This non-compliant link version is most likely causing STARS to reject ADS-B reports from these aircraft.

There is a mismatch between the ADS-B reported speed and the speed calculated using inter-report position and TOA. Fig.7. compares the ADS-B reported speed (orange line) and the calculated speed (blue markers) for aircraft A637E1. It is evident that the calculated speed exhibits large offsets from the reported speed. This observed offset in the calculated speed indicates mismatch between the reported positions and the associated TOA. This issue could be related to the non-compliant link version in the report.

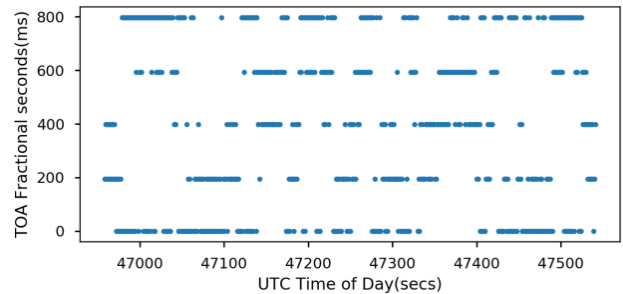

Fig. 6. Fractional portion of TOAs observed in ADS-B reports from aircraft A637E1

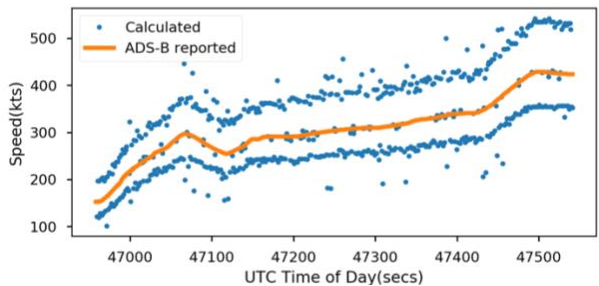

Fig. 7. Comparison of ADS-B reported speed and speed calculated using inter-report position.

## V. CONCLUSIONS

This report presents the methods of identifying and quantifying UL in ADS-B reports and the results from a latency analysis of 40 aircraft. We processed the ADS-B reports extracted from live-data recorded at the Chicago STARS site to generate these results. Using the ATPE method, UL was calculated from the ADS-B along-track position error and the reported velocity, where position errors were calculated using reference positions obtained from a pseudo-truth track. The pseudo-truth track was generated with a cubic-polynomial spline fit on track data from the 8001-LAN messages. We also implemented a point estimation method MTPES. This method produced a single latency value per track for an aircraft.

We observed that 99% of aircraft in our data set were broadcasting in non-UTC coupled mode. For these aircraft, STARS exclusively used ADS-B reports to update tracks when available. The calculated ULs for the non-UTC coupled ADS-B reports were well within the budget allowed by the DO-260B MOPS. Moreover, the mean latencies from both methods were less than ±50 ms. The latencies obtained from the MTPES method were observed to be in close agreement with the means from the ATPE method. About 1% of aircraft were broadcasting in UTC coupled mode. The reports from these aircraft were identified to have a non-compliant link version, causing STARS to reject these reports for tracking. Moreover, the reported position and TOA of these reports were found to be unsynchronized, which conflicts with the MOPS requirement for UTC coupled operation.

ACKNOWLEDGMENT

This work was supported by the Federal Aviation Administration.